\documentstyle[11pt]{article}

\newcommand{\eeq}{\end{equation}}
\newcommand{\vs}[1]{\rule[- #1 mm]{0mm}{#1 mm}}
\newcommand{\beq}{\begin{equation}}
\newcommand{\beql}{\begin{eqnarray}}
\newcommand{\eeql}{\end{eqnarray}}

\newcommand{\lam}{\lambda}

\newcommand{\eps}{\epsilon}

\newcommand{\sect}[1]{\setcounter{equation}{0}\section{#1}}
\renewcommand{\theequation}{\thesection.\arabic{equation}}

\begin{document}

\begin{titlepage}

%\rightline{\Large {April 1998}}
%\rightline{\Large {JNCASR-98-}}
%\rightline{\Large {ITP-UH-08/96}}
%\rightline{\Large {cond-mat, hep-th}}

\vs{5}

\begin{center}

{\LARGE {\bf Smoothed Correlators For Symmetric\\[.5cm]
             Double-Well Matrix Models:\\[.5cm]
             Some Puzzles and Resolutions}}\\[2cm]

{\Large E. Br\'ezin$^1$ ~~ and ~~ N. Deo$^2$}\\
{$^1$ Laboratoire~~ de~~ Physique~~ Th\'eorique}\\
{Ecole Normale Sup\'erieure}\\
{24 rue Lhomond 75231,}\\
{Paris Cedex 05, France}\\
{$^2$ Jawaharlal~~ Nehru~~ Centre} \\
{for Advanced Scientific Research}\\
{Bangalore 560064, India}\\
{Edouard.Brezin@physique.ens.fr}\\
{ndeo@jncasr.ac.in}\\

\end{center}

\vs{10}

\centerline{ {\bf Abstract}}

Some puzzles which arise in matrix models with multiple cuts are presented.
They are present in the smoothed eigenvalue correlators of these models.
First a method is described to calculate smoothed eigenvalue correlators
in random matrix models with eigenvalues distributed in a single-cut,
previous know results are reproduced. The method is extended to
symmetric two-cut random matrix models. The correlators are written in a
form suitable for application to mesoscopic systems. Connections
are made with the smooth correlators derived using the Orthogonal
Polynomial (OP) method. A few interesting observations are made
regarding even and odd density-density correlators and cross-over
correlators in $Z_2$ symmetric random matrix models.

%PACS: 05.45.+b; 05.30.-d
%Keywords: Eigenvalue distribution;
%random matrices; disconnected support

\end{titlepage}

\renewcommand{\thefootnote}{\arabic{footnote}}
\setcounter{footnote}{0}

\sect{Introduction}\label{intro}

Matrix models have been used in a wide variety of
applications, starting from quantum chaotic systems to condensed matter,
 QCD and string theory.
The recent period has seen a large increase in our understanding
of the properties of these models. In this work we have been
interested in highlighting some unusual
properties of two-cut random matrix models that have
arisen in our study. The results are unexpected as they
are not seen in matrix models when the density of eigenvalues has
a connected support. Indeed there it is  well-known \cite {AJM90, BZ93}
that the correlator is
universal, i.e. independent of the specific potential $V$ which defines the
probability measure. This is the basis for
the theory under the universality of conductance fluctuations in mesoscopic
systems \cite{B94}.
At first sight one is tempted to think that this universality persists when
the potential
is such that the support splits into disconnected segments. But it is
found that, if indeed it is again universal,
it belongs to a different universality class.
If the standard large N-limit (the random matrices are $N\times{N}$) yields
the smoothed correlation
functions up here to an arbitrary constant, different methods report
different results for this constant. Furthermore there are differences
between these correlators when the size N of the matrices is an even or an odd
integer. It is a rather intriguing phenomenon and, for instance, it is not
clear how the naive renormalization
group approach \cite{BZinn92} which consisted of integrating out one line
and one row, could deal with such situations.
We attempt here to understand and to give a unified picture of
these results.

The paper is divided as follows. It
starts by establishing the notation and conventions
and describes completely the method used for the
model with a single-cut density of eigenvalues. Previously
known results are reproduced \cite{AJM90, BZ93, B94}. Then
the method is extended to the model with two-cuts in
the density of eigenvalues, restricted to symmetric potentials.
Afterwards we devlop the formalism to include asymmetric
potentials. Here an arbitrariness
remains as the constraint on the filling factor of the two parts of the
support is not fixed at leading order in the large N-limit. The 
large N-equations for the correlator
leave us with an undetermined constant C. Previous
methods using the orthogonal
polynomials and loop equations give different results
for this constant . The orthogonal polynomial
method is briefly outlined and the resulting correlators
are sensitive to the even and oddness of the number of
eigenvalues. Further, the constant C is different for the even
and odd correlators found by the orthogonal polynomial
method and that found by the loop equations. The conclusion
summarizes these results and attempts to give an explanation
of these puzzles.

\sect{Notation, Conventions} \label{notation}

We establish the notations and conventions and develop a
method, which we extend to the two-cut model, to derive
eigenvalue correlators for random matrix models with a
single cut density of eigenvalues.

\indent
Let us work with the unitary invariant ensemble of random Hermitian
$N\times N$ matrices, with a probability distribution
\beq
P(M)={1\over Z} exp{(-N Tr V(M))}.
\eeq
Define the operator for the density of eigenvalues
\beq
\rho(x)={1\over N} \sum_{i=1}^{N} \delta (x-\lambda_i)
\eeq
and
\beq
\bar{\rho} (x) = <\rho (x)>= \int P(M) {1\over N}
Tr \delta (x-M) d^{N^2} M.
\eeq
Since $P(M)$ gives a factor
$exp( -N^2 \int V(x) \rho(x) dx)$ we have
\beq
{{\delta \bar{\rho} (x)} \over {\delta V(y)}}
= - N^2 <\rho (x) \rho (y)>_c.
\eeq
In the large N limit, we know that
\beq
\int_a^b {{\bar{\rho} (y)}\over {x-y}} dy
= {1\over 2} V^\prime (x).
\eeq
The solution is found through the averaged resolvent
\beql
G(z) &=& <{1\over{N}} Tr {1\over{z-M}}>
= \int_a^b {{\bar{\rho} (y)}\over
{z-y}} dy\nonumber\\
&=& {1\over 2} V^{\prime} (z)
-P(z) \sqrt{(z-a)(z-b)}.
\label{fz}
\eeql
Then
\beq
\pi \bar{\rho} (\lam) = P(\lam) \sqrt{(\lam-a)(b-\lam)}.
\eeq
We then need to express P as a functional of V. There are various equivalent
expressions, and in the following several will be needed. We follow here
\cite{B94} and begin by multiplying
eq. (\ref{fz}) by ${\sqrt{(z-a)(z-b)}\over z-u}$
and integrate z over a large circle C, see Fig. 1.

Since $G(z)\approx {1\over z}$ at infinity

(i).\beql \oint_c {G(z)\sqrt{(z-a)(z-b)} \over z-u}
{dz\over 2i\pi} =1 \eeql

(ii).\beql \oint_c {P(z) (\sqrt{(z-a)(z-b)})^2\over z-u} {dz\over 2i\pi}
= P(u) (u-a)(u-b)\eeql
\beql
(iii). \oint_c {V^{\prime} (z) \sqrt{(z-a)(z-b)}\over z-u}{dz\over 2i\pi}
&=&V^{\prime} (u) \sqrt{(u-a)(u-b)}\nonumber\\
&-&{1\over \pi}\int_a^b dx {{V^{\prime}(x)
\sqrt{(x-a)(b-x)}}\over {x-u}}\nonumber\\
\eeql

Therefore we obtain from eq. (\ref{fz})
\beql
1&=&{1\over 2} V^{\prime} (u) \sqrt{(u-a)(u-b)}\nonumber\\
&-& {1\over 2\pi}\int_a^b dx
{{V^{\prime} (x) \sqrt{(x-a)(b-x)}}\over x-u}
+ P(u) (u-a)(b-u).
\label{one}
\eeql
Let u approach the real axis on the cut. The integral in eq. (\ref{one})
has an imaginary part which cancels the first term of the right hand side
which is pure imaginary. The real part of eq. (\ref{one}) gives
\beq
P(\lam)(\lam-a)(b-\lam)=1+{1\over 2\pi}\int_a^b dx
{{V^{\prime} (x) \sqrt{(x-a)(b-x)}}\over x-\lam}
\eeq
i.e.
\beql
\bar{\rho}(\lam)&=& {1\over \pi} {1\over \sqrt{(\lam-a)(b-\lam)}}\nonumber\\
&&[1+{1\over 2\pi} \int_a^b dx {{V^{\prime} (x) \sqrt{(x-a)(b-x)}}\over
{x-\lam}}]\nonumber\\
\label{two}
\eeql
Now we vary V, let us first ignore the variation of a and b
(it is proved to be right below). Then
\beql
{{\delta \bar{\rho}(\lam)}\over {\delta V(\mu)}}&=&
({{\partial \bar{\rho}(\lam)}\over {\partial V(\mu)}})_{a,b}
+({{\partial \bar{\rho}(\lam)}\over {\partial a}})_{V,b}
{{\delta a}\over {\delta V(\mu)}}
+({{\partial \bar{\rho}(\lam)}\over {\partial b}})_{V,a}
{{\delta b}\over {\delta V(\mu)}}
\eeql
on the right hand side (r.h.s.) $\bar{\rho}$ is being treated
as a function of $V, a, b$ as given on the r.h.s. of eq. (\ref{two}).
We show later that
$({{\partial \bar{\rho}}\over {\partial a}})_{V,b}=0$.
[$({{\partial \bar{\rho}}\over {\partial a}})_{V,b}$
is of course not the total derivative of $\bar{\rho}$ w.r.t. a].
Then
\beql
{{\delta \bar{\rho}(\lam)}\over {\delta V(\mu)}}
&=& {1\over 2\pi^2} {1\over \sqrt{(\lam-a)(b-\lam)}}{\partial \over
\partial \mu}
{\sqrt{(\mu-a)(b-\mu)}\over {\lam-\mu}}\nonumber\\
\label{drdv}
\eeql
and one verifies easily that the result is symmetric under exchange of
$\lam$ and $\mu$
as it should. Note that the potential $V$ has disappeared from the correlator,
except indirectly through the end points $a$ and $b$ of the cut.
This universality follows here trivially from the linearity of the
$(\rho,V)$ relation.
Terms of the type
\beq
({{\delta \bar{\rho}}\over {\delta a}})_{V,b}
{{\delta a}\over{\delta V (\mu)}}
\eeq
have been ignored. The claim is that they vanish, but that's the only
(slightly) tricky part. The representation eq. (\ref{two}) is appropriate,
among several other possibilities, because if one differentiates inside
the integral with respect to a, it is still a meaningful integral. So
let us calculate
\beql
({\partial \bar{\rho} \over \partial a})_{V,b}
&=& -{1\over 2} {1\over (\lam -a)}
\bar{\rho}(\lam)+{1\over 2\pi^2} {1\over \sqrt{(\lam-a)(b-\lam)}}\nonumber\\
&+&{1\over 2} \int_a^b V^{\prime} (x) \sqrt{{b-x}\over{x-a}}
{1\over (x-\lam)} dx.
\label{rhoa}
\eeql
To prove that this is zero, let us return to eq. (\ref{fz}) and multiply
it by $\sqrt{{z-b}\over{z-a}}{1\over {z-u}}$ and integrate again over
a circle of large radius. Then
\beq
\oint_c G(z) \sqrt{{z-b}\over{z-a}}{1\over {z-u}} dz =
\oint_c {dz\over z^2}=0
\label{rhob}
\eeq
as for large $z$, $G(z)={1\over z}$,
$\sqrt{{z-b}\over {z-a}} \approx 1$
and ${1\over z-u} \approx {1\over z}$.
While the second and third terms become
\beq
-\oint_c {P(z) (z-b)\over (z-u)} {dz\over 2i\pi}=-P(u)(u-b)
\label{rhoc}
\eeq
and
\beql\label{rhod}
{1\over 2} \oint_c \sqrt{{z-b}\over {z-a}} {1\over {z-u}} V^\prime (z)
{dz\over 2i\pi} &=& {1\over 2} \sqrt{{u-b}\over{u-a}} V^\prime (u)
-{1\over 2\pi} \int_a^b \sqrt{{b-x}\over {x-a}} {V^\prime (x)\over
x-u} dx.\nonumber\\\eeql
Taking $u=\lam+i\eps$ and using ${1\over {\alpha -i\eps}}={P\over \alpha}
+i\pi \delta(\alpha)$, the integral in eq. (\ref{rhod}) has a part
which cancels the first term leaving
\beql
{1\over 2} \oint_c \sqrt{{z-b}\over {z-a}} {1\over {z-u}} V^\prime (z)
{dz\over 2i\pi} &=&
-{1\over 2\pi} \int_a^b \sqrt{{b-x}\over {x-a}} {V^\prime (x)\over
x-u} dx.
\label{rhoe}
\eeql
Repeating all the steps which led to
eq. (\ref{drdv}) i.e. combining eq. (\ref{rhob}), eq. (\ref{rhoc}),
eq. (\ref{rhoe}) one finds an expression for $\bar{\rho}$ from
\beql
{1\over 2\pi} \int_a^b \sqrt{{b-x}\over {x-a}} {V^\prime (x)\over
{x-\lam}} dx = P(\lam)(b-\lam)
\eeql
which is
\beql
{ {\bar{\rho} (\lam)} \over {(\lam-a)} } =
{1\over {2 \pi^2 \sqrt{(\lam-a)(b-\lam)}} }
{1\over 2} \int_a^b \sqrt{b-x\over x-a}
{1\over (x-\lam)} V^\prime (x) dx
\eeql
thus proving that $({\partial \bar{\rho} \over \partial a})_{V,b}=0$.
This completes the proof for the single-cut correlator.

\sect{The Double Well}\label{dw}

Now let us extend the result to eigenvalues distributed in two disjoint
bands ($[-a,-b] \cup [a,b]$). Let us first restrict ourselves to even
potentials i.e.
\beql
P(M)&=&{Z^{-1}}exp (-N Tr V(M))\nonumber\\
P(-M)&=&P(M)
\eeql
which implies for the resolvent
\beq
G(-z)=-G(z).
\eeq
Since we restrict ourselves to even potentials we cannot take a functional
derivative
of $\rho(\lam)$ with respect to an arbitrary $V(\mu)$, but we can fold the
integrations over
the positive part of the spectrum and then vary the potential.
Now
\beql
Tr V(M) &=& N\int_{-\infty}^{+\infty} d\lam \rho(\lam)V(\lam)\nonumber\\
&=& N\int_{0}^{\infty}d\lam V(\lam) [\rho(\lam)+\rho(-\lam)].
\eeql
Consequently
\beql
-{1\over N^2}{{\delta \bar{\rho} (\lam)}\over {\delta V(\mu)}}
&=& <\rho(\lam)\rho(\mu)>_c + <\rho(\lam)\rho(-\mu)>_c,
\eeql
where use has been made of
\beq
{{\delta V^\prime (x)}\over {\delta V(\mu)}}
=\delta^\prime (x-\mu).
\eeq
In the large N limit again
\beql
2 G(z) &=& V^{\prime} (z) - P(z)\sqrt{\sigma(z)}
\label{gz}
\eeql
with $\sigma(z)\equiv (z^2-a^2)(z^2-b^2)$. Note that this equation
determines uniquely $P(z), ~~a ~~\& ~~b$ ; indeed take
\beql
&& deg V = 2n\nonumber\\
&& \rightarrow
deg[P]=2n-3;\nonumber\\
\nonumber\eeql
$P(z)$ has to be odd
\beq
P(z)=\alpha_1 z + \alpha_2 z^3 + \cdots + \alpha_{n-1} z^{2n-3}
\eeq
we thus have $(n-1)+2$ unknowns. Since $G(z)\approx_{z\rightarrow \infty}
{1\over z}$ we have to fix the coefficients of eq. (\ref{gz}) at infinity
from $z^{2n-1}, z^{2n-3}, \cdots , z^{1}, z^{-1} \rightarrow
(n+1)$ conditions. Therefore no "filling" parameter creeps into the
problem (although the question of spontaneous symmetry maybe eliminated
by the assumptions here).

Now we take eq. (\ref{gz}), multiply by ${\sqrt{\sigma(z)}
\over {z(z-u)}}$
and integrate over a large circle in the z plane (using
$\sqrt{{\sigma(z)\over {z^2-u^2}}}$ also has been checked
to give the same equation), see Fig. 2. We obtain
\beql
2&=& {V^{\prime} (u) \sqrt{\sigma (u)}\over u} - {2i\over 2i\pi}
\int_b^a {{V^{\prime} (x) \sqrt{|\sigma(x)|}}\over {x(x-u)}}\nonumber\\
&+&{2i\over 2i\pi}
\int_{-a}^{-b} {{V^{\prime} (x) \sqrt{|\sigma(x)|}}\over {x(x-u)}}
-{{P(u)\sigma(u)}\over u}\eeql
which simplifies to
\beql
2+{{P(u)\sigma(u)}\over u} &=& {{V^{\prime} (u) \sqrt{|\sigma(u)|}}
\over u}\nonumber\\
&-& {1\over \pi}
\int_b^a {{V^{\prime} (x) \sqrt{|\sigma(x)|}}\over {x}}({1\over (x-u)}
+{1\over (x+u)}).
\eeql
We now let $u=\lam+i\eps$ approach the cut, say the one on the right
(it doesn't matter)
\beql
&& 2+{{P(\lam)\sigma(\lam)}\over \lam}
= {{V^{\prime} (\lam) \sqrt{|\sigma(\lam)|}}
\over \lam}\nonumber\\
&-& {1\over \pi}
\int_b^a {{V^{\prime} (x) \sqrt{|\sigma(x)|}}\over {x}}{1\over (x+\lam)}
-{1\over \pi}
\int_b^a {{V^{\prime} (x) \sqrt{|\sigma(x)|}}\over {x}}{1\over
(x-\lam-i\eps)}\nonumber\\
\eeql
In the last integral use ${1\over {\alpha-i\eps}}={PP\over \alpha}
+i\pi\delta(\alpha)$ and we obtain
\beql
2+{{P(\lam)\sigma(\lam)}\over \lam}
&=&-{1\over \pi}
\int_b^a {{V^{\prime} (x) \sqrt{|\sigma(x)|}}\over {x}}
({1\over (x-\lam)}+{1\over (x+\lam)}) dx\nonumber\\
\eeql
Let us take the derivative with respect to $V(\mu)$
($\mu$ is $>0$ by definition of V).
\beql
{\sigma(\lam)\over{\lam}}{{\delta P(\lam)} \over { \delta V(\mu)}} =
{1\over \pi} {\partial\over \partial \mu}
{\sqrt{|\sigma(\mu)|}\over \mu}
({1\over (\mu-\lam)} + {1\over (\mu+\lam)})
\label{dpdv}
\eeql
(assuming that we can show here as usual that
a counterpart of
$({{\delta \bar{\rho}} \over {\delta a}})_{V,b}
{\delta a\over \delta V(\mu)}$
and $({{\delta \bar{\rho}} \over {\delta b}})_{V,a}
{\delta b\over \delta V(\mu)}$ vanish, see Appendix A
for a proof, the exact same steps can
be followed here). Then
\beql
&&\bar{\rho} (\lam) = {1\over 2\pi}
\sqrt{|\sigma(\lam)|} P(\lam)\\
&&(\lam>0\nonumber)\\
&&{{\delta \bar{\rho}(\lam)} \over {\delta V(\mu)}}
=2 {1\over 2 \pi} {\sqrt{|\sigma(\lam)|}\over {\sigma (\lam)}}
\lam {{\partial}\over {\partial \mu}}
{\sqrt{(\mu^2-b^2)(a^2-\mu^2)}\over {\mu^2-\lam^2}}\nonumber\\
&&= - {1\over \pi} {{\lam\mu}\over
\sqrt{|\sigma(\lam)||\sigma(\mu)|}}
{1\over (\mu^2-\lam^2)^2}
[2\lam^2\mu^2-(\lam^2+\mu^2)(a^2+b^2)+2a^2b^2]\nonumber\\
\label{sdw}
\eeql
Let us check immediately the $b=0$ limit
\beq
{\lam\over \sqrt{|\sigma(\lam)|}} \rightarrow
{1\over \sqrt{(a^2-\lam^2)}}
\eeq
and the rest looks unfamiliar; but if we remember that we are computing
\beq
\rho_2(\lam,\mu)+\rho_2(\lam,-\mu)
\eeq
and
\beql
{{a^2-\lam\mu}\over {(\lam-\mu)^2}} + {{a^2+\lam\mu}\over {(\lam+\mu)^2}}
&=& -2 {[2\lam^2\mu^2-a^2(\lam^2+\mu^2)]\over (\lam^2-\mu^2)^2},
\nonumber\\
\eeql
we check that this result agrees as expected for $b=0$ with the single cut
result.
Therefore for a symmetric double well, assuming no spontaneous symmetry
breaking, we have the undisputable answer for
$\rho_2(\lam,\mu)+\rho_2(\lam,-\mu)$ i.e.  eq. (\ref{sdw}). Note that, as
expected, the short distance
behaviour of $\rho_2(\lam,\mu)$ is the same as for the single well with
only one cut.

\sect{Asymmetric Double Well}\label{adw}

In order to extract $\rho_2(\lam,\mu)$ alone we have to consider arbitrary
potentials instead of restricting ourselves to even ($Z_2$)
symmetric potentials as we have done in the above section.

Again let us start with
\beq
\rho(\lam)={1\over N} \sum_{i=1}^{N} \delta (\lam-\lam_i)
\eeq
with
\beq
\bar{\rho}(\lam)=<\rho(\lam)>.
\eeq
Since the weight contains
\beq
exp ( - N Tr V ) = exp (- N^2 \int V (\lam) \rho (\lam) d\lam)
\eeq
\beql
\rho^c_2 (\lam,\mu) = <\rho(\lam)\rho(\mu)>_c =
- {1\over N^2} {{\delta \bar{\rho} (\lam)}
\over {\delta V (\mu)}}
\eeql
\beq
\bar{\rho} (\lam) = - {1\over \pi} Im G(\lam+i0)
\eeq
\beq
G(z)= < {1\over N} Tr {1\over z-M} >
\eeq
\beq\label{E}
2 G(z) = V^{\prime} (z) - P(z) \sqrt{\sigma(z)}
\eeq
\beq
\sigma(z)=\Pi_{i=1}^{4} (z-a_i).
\eeq
See Fig. 3.

The support of the eigenvalues consists of the two segments [$a_1,a_2$] and
[$a_3,a_4$], (we assume that they are labelled by increasing order) ;
the positivity of $\rho(\lam)$ is satisfied provided the polynomial
 $P(z)$ has an odd number of zeroes between $a_2$ and $a_3$. Contrary to
the two previous cases
(\ref{E}) is not sufficient to determine fully the polynomial $P(z)$ and
the four end points of the two cuts.
Counting parameters and equations one sees readily that we miss one
parameter, which we can take as the filling
factor of one of the two wells. This factor remains undetermined at this
level of the large N-limit,
and we would have to return to a minimization of the free energy to fix it.
However, since this parameter is not fixed at this leading order, we may
ignore it and proceed as before for
finding the leading order of the correlator.

We denote
\beql
\eps_\lam &=& +1 ~~~  a_3<\lam<a_4 \nonumber\\
           && -1 ~~~  a_1<\lam<a_2
\eeql
then
\beq
\sqrt{\sigma(\lam\pm i0)} = \pm i \eps_\lam \sqrt{|\sigma (\lam)|}.
\eeq
Therefore
\beq
2\pi \rho (\lam) = \eps_\lam \sqrt{|\sigma(\lam)|} P(\lam)
\eeq
and positivity implies that $sign P(\lam)=\eps_\lam$. By
multiplication by $\sqrt{{\sigma(z)}\over {(z-u)(z-v)}}$ integration
over a large circle, we obtain
\beql
2+ {{P(u) \sigma (u) - P(v) \sigma (v)}\over {u-v}}
&=& { {V^{\prime} (u) \sqrt{\sigma (u)} - V^{\prime} (v)
\sqrt{\sigma (v)}} \over {u-v} }\nonumber\\
&-&{1\over \pi} (\int_{a_3}^{a_4}-\int_{a_1}^{a_2}) {dx V^{\prime} (x)
\sqrt{|\sigma (x)|}
\over {(x-u)(x-v)}}.
\eeql
Let $u=\lam+i\eps$ and $v=\mu+i\eta$ then
\beql
2+{{P(\lam)\sigma(\lam)-P(\mu)\sigma(\mu)} \over {\lam - \mu}}
&=& {1\over \pi} (\int_{a_1}^{a_2}-\int_{a_3}^{a_4}){V^\prime (x)
\sqrt{|\sigma(x)|}
\over (x-\lam)(x-\mu)}\\
R(\lam,\nu)&=&{\delta P(\lam)\over \delta V(\nu)}\\
{{\sigma(\lam)R(\lam,\nu)-\sigma(\mu)R(\mu,\nu)}\over {\lam-\mu}}
&=& \eps_\nu {1\over \pi} {\partial\over \partial \nu}
{\sqrt{|\sigma(\nu)|}\over {(\nu-\lam)(\nu-\mu)}}
\label{rhomunu}
\eeql
(assuming once more an equivalent form of ${{\delta \bar{\rho}}
\over {\delta a_i}}{{\delta a_i}\over {\delta V}}$ to be zero,
see Appendix A). Hence
\beql
\sigma(\lam)R(\lam,\nu)-{\eps_\nu\over \pi} {\partial\over \partial
\nu} {\sqrt{|\sigma(\nu)|}\over \nu-\lam}
&=& \sigma (\mu) R (\mu,\nu) - {{\eps_\nu} \over \pi}
{\partial \over \partial \nu} {\sqrt{|\sigma (\nu)|}\over {\nu-\mu}}.
\eeql
From these equations one finds
\beq
\sigma (\lam) R (\lam, \nu) - {\eps_\nu \over \pi} {\partial \over
\partial \nu} {\sqrt{|\sigma (\nu)|}\over {\nu-\lam}}
={h_{\epsilon_{\nu}}(\nu)\over \sqrt{|\sigma (\nu)|}}
\label{A}\eeq
and we are left with two unknown functions $h_+$ and $h_-$ of a single
variable.
This gives the connected correlator
\beq
\rho^c_2 (\lam,\mu) = - {1\over{2{\pi}{N^2}}}  \eps_\lam
\sqrt{|\sigma(\lam)|} R(\lam,\mu),
\label{C}\eeq
i.e.
\beq
2\pi N^2 \rho^c_2 (\lam,\mu) = {\eps_\lam\over \sqrt{|\sigma (\lam)|}}
\{ {\eps_\mu \over \pi} {{\partial} \over {\partial \mu}}
{\sqrt{|\sigma(\mu)|}\over {\mu-\lam}} +{h_{\epsilon_{\mu}}(\mu)\over
{\sqrt{|\sigma(\mu)|}}} \}.
\label{B}
\eeq
From its definition the two-point correlator is symmetric under exchange of
the two eigenvalues
\beq
\rho^c_2 (\lam,\mu) = \rho^c_2 (\mu,\lam).
\label{C}
\eeq
This imposes the following constraints:
\beq h_{+}(\mu) +  h_{-}(\mu) = 0 \eeq
and
\beq \pi[ h_{+}(\mu) -  h_{+}(\lam) ] = \sqrt{|\sigma (\lam)|}{{\partial}
\over {\partial \lam}}{\sqrt{|\sigma(\lam)|}\over {\lam-\mu}}
-\sqrt{|\sigma (\mu)|}{{\partial} \over {\partial
\mu}}{\sqrt{|\sigma(\mu)|}\over {\mu-\lam}}.\eeq
A straightforward algebra gives from there the function $h_+$ up to an
arbitrary constant:

\beq h_{+}(\lam) = {1\over{\pi}}(\lam^2 - {1\over{2}} s \lam + C)\eeq
with \beq
s = a_1 + a_2 + a_3 +a_4. \eeq
We are thus left with one undetermined constant in the two-point function:
\beql
4{\pi}^2 N^2 \rho^c_2 (\lam,\mu) &=& {{{\eps_\lam \eps_\mu}\over
{\sqrt{|\sigma(\lam)|}\sqrt{|\sigma(\mu)|}}}({{\sigma(\lam) +
\sigma(\mu)}\over{(\lam-\mu)^2}}}\nonumber\\ &+&  {{\sigma'(\lam)
-\sigma'(\mu)}\over{(\lam-\mu)}} + \lam^2 + \mu^2 -{s\over{2}}(\lam+\mu) +
2C).\nonumber\\
\label{rholamnu}
\eeql

Let us verify that, without any restriction on the constant $C$, this
result satisfies the normalisation condition
\beq
\int d\nu \rho^c_2 (\lam,\nu) =0,
\eeq
which follows from the definition of $\rho^c_2$.
Returning to (\ref{C})
\beq
\int d\mu {\partial\over \partial \mu} {\sqrt{\sigma(\mu)}
\over {\mu-\lam}}
=\int_{a_1}^{a_2} d\mu {\partial\over \partial \mu} {\sqrt{\sigma(\mu)}
\over {\mu-\lam}}
+\int_{a_3}^{a_4} d\mu {\partial\over \partial \mu} {\sqrt{\sigma(\mu)}
\over {\mu-\lam}} =0
\eeq
since $\sigma$ vanishes at the end points. (This point is in fact slightly
delicate, since
there is a non-integrable singularity at $\mu=\lam$. In the literature
concerning the application of random matrices to
the calculation of the fluctuations of the conductance in mesoscopic
systems \cite{B94}, this integration throughout the singularity
is done in a routine way. A proper justification of the procedure implies
returning to the true
correlation function, before the smoothing which produces spurious
short-distance singularities through replacements such as
(${\sin^2 x\over x^2} \rightarrow {1\over 2x^2}$). The smoothing is
produced here by the large N limit.)
Next consider
\beq
\oint dz {(z^2-sz/2)\over \sqrt{\sigma(z)}}
\eeq
over a large circle. Note that $\sqrt{\sigma(z)}=z^2[1-{s\over 2z}
+O({1\over z^2})]$. Consequently ${{z^2-sz/2}\over \sqrt{\sigma (z)}}
=1+O({1\over z^2})$. Since there is no coefficient of $1\over z$ the
integral vanishes. Shrinking the contour over the cuts we obtain
\beq
(\int_{a_1}^{a_2}-\int_{a_3}^{a_4}) d\nu {{\nu^2-s\nu/2}
\over \sqrt{|\sigma(\nu)|}}=0.
\eeq
Therefore
\beql
2\pi N^2 (\int_{a_1}^{a_2} d\mu \rho_2^c +\int_{a_3}^{a_4} d\mu \rho_2^c)
&=&C{\eps_\lam\over {\sqrt{|\sigma (\lam)|}}}
[ \int_{a_3}^{a_4}
{d\mu \over \sqrt{|\sigma(\mu)|}}\nonumber\\
&-&\int_{a_1}^{a_2}
{d\mu \over \sqrt{|\sigma(\mu)|}} ].\nonumber\\
\eeql
Again taking a very large circle
\beq
\oint {dz\over \sqrt{\sigma(z)}}=0
\eeq
since the coefficient of $1\over z$ vanishes. Therefore, shrinking
the circle,
\beq
\int_{a_3}^{a_4} {d\mu \over \sqrt{|\sigma(\mu)|}}=\int_{a_1}^{a_2}
{d\mu \over \sqrt{|\sigma(\mu)|}}
\eeq
which shows that the normalization is correct for any value of $C$.

Let us specialize to the symmetric double-well ($a_1=-b$, $a_2=-a$,
$a_3=a$ and $a_4=b$)
\beq
|\sigma(\lam)|=(\lam^2-a^2)(b^2-\lam^2).
\eeq
After a few lines
\beq
2\pi^2\rho^c_2 (\lam,\mu)={\eps_\lam \eps_\mu
\over {\sqrt{|\sigma(\lam)||\sigma(\mu)|}}}
{1\over (\mu-\lam)^2} [ C(\mu-\lam)^2
+\lam\mu (\lam\mu-a^2-b^2)+a^2 b^2 ].
\label{rhosdw}
\eeq

Several remarks
(i). The result is manifestly symmetric under
$\lam \leftrightarrow \nu$
(ii). C remains an unknown function of a \& b
(iii). If we assume that C vanishes when b=0 then
since
\beq
\lim_{a\rightarrow 0} \sqrt{|\sigma(\lam)|}
=\eps_\lam \lam (a^2-\lam^2)
\eeq
we recover the single band result.
(iv). The cross-correlator $\rho_c^2 (\lam,-\nu)$ is simply
eq. (\ref{rhosdw}) with $\nu$ replaced by $-\nu$. Combining
eq. (\ref{rhosdw}) with $\rho_c^2 (\lam,-\nu)$ we reproduce the
result eq. (\ref{sdw}) in section \ref{dw}.
(v). We also note that for
$C=-{1\over 2} ((a^2+b^2)-(a+b)^2 {E(k)\over K(k)})$ we recover
the connected density-density correlator derived from
the connected Green's function (where $E(k)$ and $K(k)$
are complete elliptic integrals of first and second kind 
and $k={2\sqrt{ab}\over {(a+b)}}$) in eq. (15) of ref. \cite{AA96}.

\sect{Orthogonal Polynomials, The
Kernel, Odd and Even N}\label{OP}

This section follows the notation of ref. \cite{D97}.
Let us calculate the Kernel $K_N (\mu,\nu)$ for $N$ even and $N$ odd
\beq
K_N (\mu,\nu) = {\sqrt{R_N}\over N} [{ {\psi_N (\mu) \psi_{N-1} (\nu)
-\psi_{N-1} (\mu) \psi_N (\nu)}\over {(\mu-\nu)} }].
\eeq
The asymptotic ansatz for $\psi_n (\lam)$ for $N\rightarrow \infty$
but $N-n$ finite is
\beql
\psi_n (\lam) &=& {1\over \sqrt{f(\lam)}} [ \cos (N\zeta-(N-n)\phi
+ \chi + (-1)^n \eta) (\lam) + O({1\over N}) ] \nonumber\\
f(\lam) &=& {\pi\over 2\lam} ({{b^2-a^2}\over 2}) \sin 2\phi (\lam)
\nonumber\\
\zeta^\prime (\lam) &=& -\pi \rho (\lam)\nonumber\\
\cos 2 \phi (\lam) &=& {{\lam^2-(a^2+b^2)/2}\over {(b^2-a^2)/2}}
\nonumber\\
\cos 2\eta (\lam) &=& {{b\cos\phi(\lam)}\over \lam}\nonumber\\
\sin 2\eta (\lam) &=& {{a\sin\phi(\lam)}\over \lam}.\nonumber\\
\eeql
Substituting
\beql
\psi_N (\lam) &=& {1\over \sqrt{f(\lam)}} [ \cos (N\zeta-(N-N)\phi+\chi
+(-1)^N \eta) (\lam)  ] \\
\psi_{N-1} (\lam) &=& {1\over \sqrt{f(\lam)}} [ \cos (N\zeta
-(N-N+1)\phi+\chi+(-1)^{(N-1)} \eta) (\lam)  ] \nonumber\\
\eeql
we get
\beql
K_N (\mu,\nu)&=&{\sqrt{R_N}\over {N(\mu-\nu)\sqrt{f(\mu)f(\nu)}}}
\cos N h(\mu) \cos N h(\nu)\nonumber\\
&\times&( \cos \phi (\nu) \cos 2 \eta (\nu)
-(-1)^N \sin \phi (\nu) \sin 2 \eta (\nu)\nonumber\\
&-& \cos \phi (\mu) \cos 2 \eta (\mu)
+ (-1)^N \sin \phi (\mu) \sin 2 \eta (\mu))\nonumber\\
&+& \sin N h (\nu) \cos N h (\mu) (\sin \phi (\nu) \cos 2 \eta (\nu)
+ (-1)^N \sin 2 \eta (\nu) \cos \phi (\nu))\nonumber\\
&-& \sin N h (\mu) \cos N  h (\nu) (\sin \phi (\mu) \cos 2 \eta (\mu)
+ (-1)^N \sin 2 \eta (\mu) \cos \phi (\mu))\nonumber\\
\eeql
where
\beq
N h(\mu)=(N \zeta + \chi + (-1)^N \eta) (\mu).
\eeq
On simplifying further
\beql
K_N (\mu,\nu) &=&
{\sqrt{R_N}\over {2N(\mu-\nu)\sqrt{f(\mu)f(\nu)}}}\nonumber\\
&&[ (\cos (N h(\mu) + N h(\nu)) + \cos (N h(\mu) - N h(\nu)))\nonumber\\
&&\times({b\over \nu} \cos^2 \phi (\nu)
- (-1)^N {a\over \nu} \sin^2 \phi (\nu)
-{b\over \mu} \cos^2 \phi (\mu) + (-1)^N {a\over \mu} \sin^2 \phi
(\mu))\nonumber\\
&+&(\sin (N h (\nu) + N h (\mu)) + \sin (N h (\nu)- N h (\mu)))\nonumber\\
&&\times({b\over \nu} + (-1)^N {a\over \nu}) \sin \phi (\nu) \cos \phi
(\nu)\nonumber\\
&-&(\sin (N h (\mu) + N h (\nu)) + \sin ( N h (\mu) - N h (\nu)))\nonumber\\
&&\times({b\over \mu} + (-1)^N {a\over \mu}) \sin \phi (\mu) \cos \phi (\mu)].
\eeql
Squaring and averaging we get after some tedious algebra
\beql
<K_N^2 (\mu,\nu)> &=& {R_N\over {4N^2(\mu-\nu)^2 f (\mu) f (\nu)
2 \nu \mu (b^2-a^2)^2/4}}\nonumber \\
&&\times[ 2 \nu \mu {(b^2-a^2)^2\over 2}
- \nu^2 \mu^2 (2 ab (-1)^N + a^2 + b^2)\nonumber\\
&+& {(a^2+b^2)\over 2} (\nu^2+\mu^2) (2 ab (-1)^N + a^2 + b^2)\nonumber\\
&-& 2 a b (-1)^N a^2 b^2 - (a^2+b^2) {(b^4+a^4)\over 2}\nonumber\\
&-& (\nu^2+\mu^2) {(b^2-a^2)^2\over 2}
+ {2\over 4} (b^2-a^2)^2 (a^2+b^2) ].
\eeql
Simplifying for N even,
\beql
<K_N (\mu,\nu)^2> &=& {{R_N (a+b)^2}\over {4 N^2 (\mu-\nu)^2 f (\mu)
f (\nu) \nu \mu {(b^2-a^2)\over 2}} }\nonumber\\
&&\times[\nu \mu (b^2+a^2) - \nu^2 \mu^2 - a^2 b^2 + (\nu-\mu)^2 ab
]\nonumber\\
\eeql
while for N odd
\beql
<K_N (\mu,\nu)^2> &=& {{R_N (a-b)^2}\over {4 N^2 (\mu-\nu)^2 f (\mu)
f (\nu) \nu \mu {(b^2-a^2)\over 2}} }\nonumber\\
&&\times[\nu \mu (b^2+a^2) - \nu^2 \mu^2 - a^2 b^2 - (\nu-\mu)^2 ab
].\nonumber\\
\eeql
Note that $R_{N_{even}} (a+b)^2 = A (a+b)^2 = {(a-b)^2\over 4} (a+b)^2
= {(a^2-b^2)^2\over 4}$ and $R_{N_{odd}} (a-b)^2 = B (a-b)^2
= {(a^2-b^2)^2\over 4}$.
Comparing this expression with that found by the previous method of
section \ref{adw}, we find that $C=(-1)^N ab$.
The standard large N-limit techniques of analyzing matrix models
like the loop equation method ref. \cite{AJM90, AA96} and renormalization
group ref. \cite{BZinn92} assume a smooth behavior with respect to N
at large N. The result that C differs for odd or even N by terms of order
one suggests that these methods may need to be revisited in the context
of random matrix models with eigenvalue distributions with gaps.

\sect{Conclusion}

To conclude we have outlined a method which reproduces known results for
the single cut model and extended it to the two-cut random matrix model. 
The two-point density-density correlator contains
a derivative part familiar from the
single cut model but in addition contains a non-trivial non-derivative
piece. It is further seen that different methods give different values
for the two-point correlator. The orthogonal polynomial method is
briefly outlined and gives different values for the non-derivative piece
for even and odd eigenvalues. The loop equation method gives a different
result. The difference in the results are in the non-derivative part of the
two-point density-density correlator. The method outlined unifies these
differences in a constant C which takes different values. Different values
of C found from the orthogonal polynomial and loop equation methods are
identified.

This raises several questions regarding the analysis of this model.
One possibility is that the even-odd differences may require some
care in handling the large N techniques of random matrix models
e.g. loop equations and renormalization group. Another question
relates to spontaneous breaking of the $Z_2$ symmetry in the large
N limit. In this context, for the $Z_2$ symmetric random matrix models
with two wells an infinite family of solutions which break the $Z_2$
symmetry and have the same free energy as the $Z_2$ symmetric solution
but different connected correlators have been identified in
ref. \cite{BDJT93}. It would be interesting to compare whether the
different solutions noted here correspond to some of the multiple
solutions of ref. \cite{BDJT93}.
Finally let us note that when the number of connected components for the
support of the eigenvalues
changes, one finds a new universality class for the correlators. It is
thus not completely obvious that
it is legitimate to use the simple one cut function in the application to
mesoscopic fluctuations. It seems interesting to us that this simple
system, namely N charges confined by a symmetric double-well with a
logarithmic repulsion between the charges, exhibits such rich behavior.

\vskip 5mm

\begin{flushleft}
\underline{Acknowledgements}:\hfill\break
EB would like to thank Ivan Kostov for discussions.
ND thanks the Abdus Salam ICTP, Trieste, Italy; Ecole Normale Sup\'erieure,
Paris, France; Isaac Newton Institute, Cambridge, England;
and NEC, Princeton, USA; for support and
hospitality where part of this work was done. 
Special thanks goes to the Raman Research Institute 
for support and encouragement. Thanks also to C. Dasgupta,
S. Jain, H. R. Krishnamurthy, N. Kumar, R. Nityananda,
N. Mukunda, T. V. Ramakrishnan, C. N. R. Rao and B. S. Shastry
for encouragement and discussions.
\end{flushleft}

\vskip 5mm

\setcounter{equation}{0}
\renewcommand{\theequation}{A.\arabic{equation}}
\noindent
{\bf Appendix A}

\vskip 5mm

Eq. (\ref{rhomunu}) was derived under the assumption that a
counterpart of $({{\delta \bar{\rho}}\over {\delta a_i}})_{V,b_i}=0$.
Here we prove this result for the asymmetric double-well.
(Following a similar procedure we can show that an equivalent form of
$({{\delta \bar{\rho}}\over {\delta a_i}})_{V,b_i}=0$
from which eq. (\ref{dpdv}) follows for the symmetric double well). From
eq. (\ref{rhomunu}) it is easy to see that we have to prove the following
equation equivalent to
$({{\delta \bar{\rho}}\over {\delta a_i}})_{V,b_i}=0$
in the single well problem
\beql
&&{{(-2\pi \bar{\rho}(\lam) \eps_\lam
{{\delta\sqrt{|\sigma(\lam)|}}\over  {\delta_a}}
+ 2\pi \bar{\rho}(\mu) \eps_\mu
{ {\delta \sqrt{|\sigma(\mu)|}} \over {\delta a} })}\over
{(\lam-\mu)} }\\
&&= {1\over \pi} \int_a^b
{ { V^\prime (x) { {\delta \sqrt{|\sigma(x)|}} \over {\delta a}} }
\over {(x-\lam)(x-\mu)} }
- {1\over \pi} \int_c^d
{ { V^\prime (x) { {\delta \sqrt{|\sigma(x)|}} \over {\delta a}} }
\over {(x-\lam)(x-\mu)}}.
\label{rholam}
\eeql
Let us take
\beq
2G(z) = V^\prime (z) - P(z) \sqrt{(z-a)(z-b)(z-c)(z-d)}.
\eeq
Multiple by ${1\over {(z-u)(z-v)}}{\sqrt{\sigma(z)}\over (z-a)}$
and integrate over a large circle. The first term
\beql
2 \oint_c {G(z)\over {(z-u)(z-v)}}
{\sqrt{\sigma(z)}\over (z-a)} {dz\over
{2\pi i}}=\oint_c {1\over z^2} dz=0
\eeql
as for large $z$ $G(z)\approx {1\over z}$, $\sqrt{\sigma(z)}\approx z^2$,
${1\over {(z-u)(z-v)}}\approx {1\over z^2}$ and ${1\over z-a}\approx
{1\over z}$. The third term becomes
\beql
&&\oint_c {{P(z)\sigma(z)}\over {(z-u)(z-v)(z-a)}} {dz\over 2\pi i}
\nonumber\\
&&={{P(u)(u-b)(u-c)(u-d)-P(v)(v-b)(v-c)(v-d)}\over {(u-v)}}.\nonumber\\
\eeql
While the second term is
\beql
\oint_c {V^\prime (z)\over {(z-u)(z-v)}} {\sqrt{\sigma(z)}\over {(z-a)}}
{dz\over 2\pi i}={{\sqrt{\sigma(u)} V^\prime (u)}\over {(u-a)(u-v)}}
+{{\sqrt{\sigma(v)} V^\prime (v)}\over {(v-a)(v-u)}}\nonumber\\
+{1\over \pi i} \int_a^b {V^\prime (x) \over {(x-u)(x-v)}}
{\sqrt{\sigma(x)}\over {(x-a)}}
+{1\over \pi i} \int_c^d {V^\prime (x) \over {(x-u)(x-v)}}
{\sqrt{\sigma(x)}\over {(x-a)}}.
\eeql
On using $u=\lam+i\eps$ and $v=\mu+i\eps$ with $\lam$, $\mu$ 
on the right hand cut, ${1\over {x-\lam -i\eps}}=
{P\over x-\lam}+i\pi \delta(x-\lam)$ and $\sqrt{\sigma(x)}=i\eps_\lam
|\sqrt{\sigma(x)}|$ the second integral simplifies to
\beql
\oint_c {V^\prime (z)\over {(z-u)(z-v)}} {\sqrt{\sigma(x)}\over {(z-a)}}
{dz\over 2\pi i} &=& {1\over \pi} \int_a^b {{V^\prime (x) |\sqrt{\sigma(x)}|
dx}\over {(x-\lam)(x-\mu)(x-a)}}\nonumber\\
&-& {1\over \pi} \int_c^d
{{V^\prime (x) |\sqrt{\sigma(x)}| dx} \over {(x-\lam)(x-\mu)(x-a)}}.
\eeql
Combining these three terms and simplifying we get eq. (\ref{rholam})
which is what is needed in order to get eq. (\ref{rhomunu}).

\vskip 5mm

%\bibliography{herm}
%\bibliographystyle{unsrt}

\newcommand{\NP}[3]{{\it Nucl. Phys. }{\bf B#1} (#2) #3}
\newcommand{\PL}[3]{{\it Phys. Lett. }{\bf B#1} (#2) #3}
\newcommand{\PR}[3]{{\it Phys. Rev. }{\bf #1} (#2) #3}
\newcommand{\PRL}[3]{{\it Phys. Rev. Lett. }{\bf #1} (#2) #3}
\newcommand{\IMP}[3]{{\it Int. J. Mod. Phys }{\bf #1} (#2) #3}
\newcommand{\MPL}[3]{{\it Mod. Phys. Lett. }{\bf #1} (#2) #3}
\newcommand{\JP}[3]{{\it J. Phys. }{\bf A#1} (#2) #3}

\newpage

\noindent
Figure captions:
\bigskip

\noindent
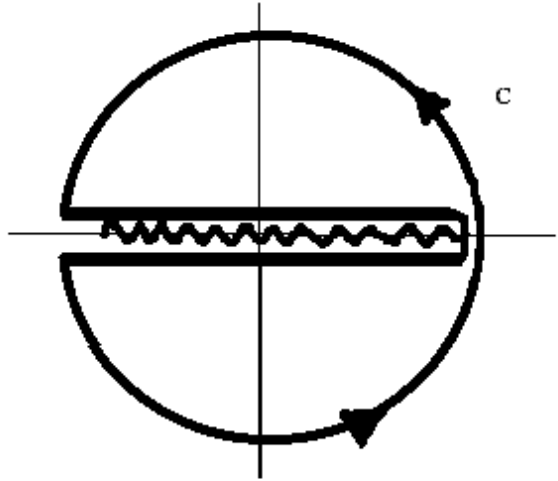
Fig. 1. The complex z-plane with one-cut and contour
used for evaluating the two-point density-density
correlator for the one-cut random matrix model.
\bigskip

\noindent
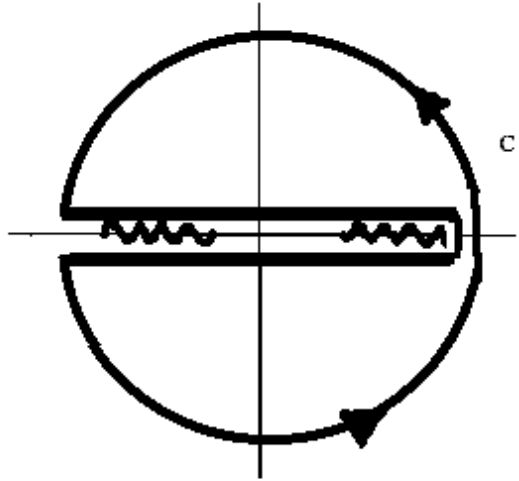
Fig. 2. The complex z-plane with two-cuts and contour
used for evaluating the two-point density-density
correlator for the two-cut random matrix model.
\bigskip

\noindent
Fig. 3. The complex z-plane with asymmetric two-cuts
and contour used for evaluating the two-point density-density
correlator for the asymmetric two-cut random matrix model.
\bigskip

\end{document}